\documentclass[aps,floatfix,twocolumn]{revtex4}
\usepackage{graphics,epsfig}
\usepackage{graphicx}

\begin{document}

\title{Galactic Halos are Einstein Clusters of WIMPs}
\author{Kayll Lake\cite{email}}
\affiliation{Department of Physics, Queen's University, Kingston,
Ontario, Canada, K7L 3N6 }
\date{\today}

\begin{abstract}
It is shown that an Einstein cluster of WIMPs, WIMPs on stable
circular geodesic orbits generating the static spherically
symmetric gravitational field of a galactic halo, can exactly
reproduce the rotation curve of any galaxy simply by adjusting the
local angular momentum distribution and consequent number
distribution of the WIMPs. No new physics is involved (assuming
the forthcoming discovery of WIMPs). Further, stability of the
orbits can require an inner truncation of the halo and an explicit
example of this is given. There is no exact Newtonian counterpart
to the model presented since pure Newtonian gravity fails to
account for the contribution made by the angular momentum
distribution of the particles in creating the gravitational field.
In effect, a galactic halo is supported by the hoop stresses
created by the orbiting WIMPs.
\end{abstract}
\maketitle

Many years ago Einstein \cite{einstein} showed that one can
introduce rotation without global angular momentum. He used a
cluster of randomly distributed gravitating dust (non-colliding)
particles each on a circular geodesic orbit about the center of
symmetry of the gravitational field produced by the entire
distribution of particles. In this letter I show that this model,
in a modern context, can solve all dynamical issues associated
with galactic rotation curves, one of the foremost problems in all
of physical science today. It is proposed that the dark matter
halos are the gravitational fields produced by massive particles
(WIMPs) in orbit about galactic centers \cite{dark}. The
gravitational contribution of the visible components is, in
excellent approximation, simply ignored here.

Einstein's heuristic arguments can be made more precise \cite
{heuristic} and here we simply sketch the construction of the
energy-momentum tensor. Consider a static spherically symmetric
continuous distribution of particles (WIMPs) of (rest) mass
$m\neq0$ and 4 - momenta $p_{\alpha}$ which are on stable circular
geodesic orbits about the center of symmetry. (The existence and
stability of these orbits is discussed below.) In terms of
``curvature" coordinates the template for our calculations takes
the form \cite{metric}
\begin{equation}
ds^2=\frac{dr^2}{1-\frac{2M(r)}{r}}+r^2d\Omega^2-e^{2\Phi(r)}dt^2,
\label{standardform}
\end{equation}
where $d\Omega^2$ is the metric of a unit sphere ($d
\theta^2+\sin^2\theta d \phi^2$). Now let $n$ be the local number
density of particles and at some spatial point $\mathcal{P}$
(labelled, say, by
$(r_{_{\mathcal{P}}},\theta_{_{\mathcal{P}}},\phi_{_{\mathcal{P}}})$)
construct
\begin{equation}
T_{\alpha}^{\beta}|_{_{\mathcal{P}}}\equiv\frac{n}{m}<p_{\alpha}\;p^{\beta}>
\label{energymomentum}
\end{equation}
where the average $<\;>$ is taken at $\mathcal{P}$ and over all
particles passing through $\mathcal{P}$. The total angular
momentum $L$ (defined as usual by
$L^2=p_{\theta}^2+p_{\phi}^2/\sin^2\theta \neq 0 $) is of course
constant along each trajectory but we also assume that it is
constant along every trajectory through $\mathcal{P}$ thus
arriving at the main results \cite{einstein}
\begin{equation}
<p_{\theta}^{}>=<p_{\phi}^{}>=<p_{\theta}^{}\;p_{\phi}^{}>=0
\label{avg1}
\end{equation}
and
\begin{equation}
<p_{\theta}^2> =\frac{<p_{\phi}^2>}{\sin^2
\theta_{_{\mathcal{P}}}}=\frac{L^2}{2}.\label{avg2}
\end{equation}
Moving to a continuous distribution and maintaining spherical
symmetry we have $L=L(r)$ and $n=n(r)$ so that finally
\begin{equation}
T_{t}^{t}=-mn(r)(1+\mathcal{L}^2(r))\label{tt}
\end{equation}
and
\begin{equation}
T_{\theta}^{\theta}=T_{\phi}^{\phi}=\frac{mn(r)}{2}\mathcal{L}^2(r)\label{thetatheta}
\end{equation}
where
\begin{equation}
\mathcal{L}(r)\equiv\frac{L(r)}{mr}.\label{L}
\end{equation}
All other components of $T_{\alpha}^{\beta}$ vanish and
$\nabla_{\beta}T^{\beta}_{\alpha}$ vanishes identically.

Let us now make use of the standard Einstein equations
$G_{\alpha}^{\beta}=8 \pi T_{\alpha}^{\beta}$ in the coordinates
defined by (\ref{standardform}) and using the energy-momentum
tensor derived above. First, since $T_{r}^{r}=0$, we can solve for
$M(r)$ algebraically from $G_{r}^{r}=0$ to obtain
\begin{equation}
M(r)=\frac{r^2 \Phi^{'}}{1+2r\Phi^{'}}\label{M(r)}
\end{equation}
where $^{'}\equiv d/dr$. Next, writing
$-T_{t}^{t}/T_{\theta}^{\theta}=-G_{t}^{t}/G_{\theta}^{\theta}$ we
arrive at
\begin{equation}
r\Phi^{'}=\frac{\mathcal{L}^2(r)}{1+\mathcal{L}^2(r)}\label{Phiprime}
\end{equation}
so that (\ref{M(r)}) can be written in the equivalent form
\begin{equation}
M(r)=\frac{r
\mathcal{L}^2(r)}{1+3\mathcal{L}^2(r)}.\label{M(r)new}
\end{equation}
Finally, from (\ref{tt}) and (\ref{thetatheta}) since
$T_{t}^{t}+2T_{\theta}^{\theta}=-mn$, we have
\begin{equation}
4 \pi mn=\frac{(1-r\Phi^{'})(r
\Phi^{''}+2\Phi^{'}+2r(\Phi^{'})^2)}{r(1+2r
\Phi^{'})^2},\label{mn}
\end{equation}
which, with the aide of (\ref{Phiprime}), can also be given in the
form
\begin{equation}
4 \pi
mn=\frac{\mathcal{L}(2r\mathcal{L}^{'}+\mathcal{L}(1+3\mathcal{L}^2))}{r^2(1+\mathcal{L}^2)(1+3\mathcal{L}^2)^2}.\label{mnl}
\end{equation}
Of course, $mn$ has to be $\geq0$. In this model there is no way
to determine $m$ and $n$ separately (an advantage discussed
below). We refer to the product $nm$ as the ``number
distribution". As (\ref{mnl}) makes clear, in this model the
number distribution $mn$ can be considered a consequence of the
angular momentum distribution $\mathcal{L}$.

If we were given $\mathcal{L}(r)$ then the gravitational field of
a halo would be complete up to quadrature in (\ref{Phiprime}). As
discussed in detail previously \cite{lake}, it is $\Phi(r)$ (and
not $\mathcal{L}(r)$) which follows directly from observations of
galactic rotation curves. However, important constraints can be
placed on $\mathcal{L}(r)$. In the notation of \cite{lake}, the
existence of these circular orbits requires $0<r\Phi^{'}<1$ which
with (\ref{Phiprime}) translates simply to
$0<\mathcal{L}(r)^2<\infty$ but shows that the first term in
(\ref{mn}) is positive. The stability of the circular geodesic
orbits requires $3\Phi^{'}+r\Phi^{''}>2r(\Phi^{'})^{2}$ which
gives $r\mathcal{L}\mathcal{L}^{'}>-\mathcal{L}^2$ and shows that
the second term in (\ref{mn}) is positive only for $r\Phi^{'}>1/4$
($\mathcal{L}^2>1/3$). A mapping onto the observer's plane
requires $\Phi^{'}>r\Phi^{''}+2r(\Phi^{'})^2$ so that
$r\mathcal{L}\mathcal{L}^{'}<\mathcal{L}^2$ and so the
distribution of angular momentum should satisfy
\begin{equation}
-\mathcal{L}^2<r\mathcal{L}\mathcal{L}^{'}<\mathcal{L}^2\label{Llimits}
\end{equation}
which, when applied to (\ref{mnl}) directly repeats the important
result: The stability of the circular geodesics orbits requires
\begin{equation}
\mathcal{L}^2 > \frac{1}{3}.\label{stability}
\end{equation}
Let $r_{0}$ signify the border of stable orbits defined by
$\mathcal{L}^2(r_{0})=1/3$. If $\mathcal{L}^2(r)$ is monotone
increasing then the region $r< r_{0}$ must be excluded and the
halo distribution is necessarily a thick shell. An explicit
example of this is shown below. (If one was to measure the
``relativistic" nature of the configurations by way of (say)
$2M(r)/r$ then they are always highly ``relativistic", starting
with $r \sim 6 M(r)$ and becoming more ``relativistic" ($r \sim 3
M(r)$) as $\mathcal{L}^2 \rightarrow\infty$. This notion of
``relativistic" is, however, based on the Schwarzschild vacuum
($M=$ constant) and it is inappropriate here. This is discussed
further below.)

 As discussed previously \cite{lake}, observations of galactic
rotation curves are reported by way of the ``optical convention"
$v \equiv (\lambda_{o}-\lambda_{e})/\lambda_{e}$ so that
\begin{equation}
\frac {\Phi^{'}b^2}{r(1-r\Phi^{'})}=(v(b)-v(b=0))^2
\label{equivalence}
\end{equation}
where the mapping from the observer's plane to $r$ is given by way
of the classical impact parameter
\begin{equation}
b^2=\frac{r^2}{e^{2\Phi}}. \label{map}
\end{equation}
Observations therefore give $\Phi$ up to quadrature and as a
consequence the full geometry, the angular momentum distribution
$\mathcal{L}$ and consequently the number distribution $mn$. In
any particular case one can take the view that the angular
momentum distribution of the orbiting WIMPs has been adjusted to
exactly match the observed rotation curve. Equivalently, the
observed rotation curve is a consequence of the distribution
$\mathcal{L}$.

\begin{figure}[ht]
\epsfig{file=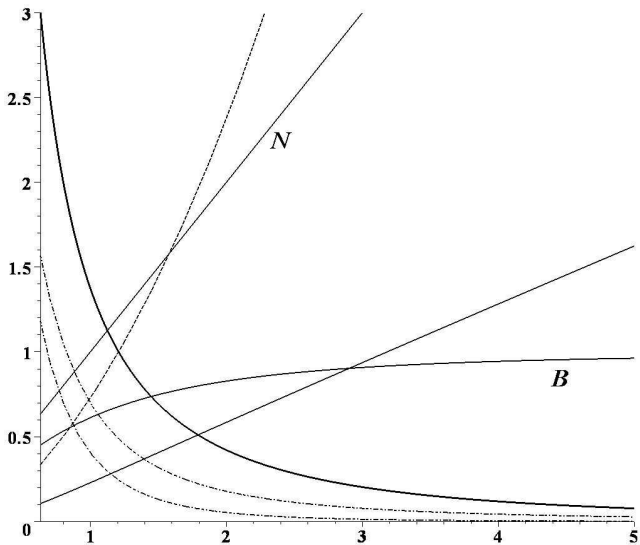,height=3in,width=3.5in,angle=0}
\caption{\label{wmodelprop}Some representative properties of the
model which follows from (\ref{ps}). For this figure
$\alpha=\beta=1$ and $c=2$. These values are for demonstration
purposes only. The thick solid curve gives $\mathcal{W}(y)$, and
the thin solid curve gives $M(r)$ which is essentially linear. The
dashed increasing curve gives $\mathcal{L}^2$ and the lower
dashdot decreasing curve gives $8\pi mn(r)$ which lies below  the
effective density, the dashdot decreasing curve given by $8\pi
mn(r)(1+\mathcal{L}^2)$ (in these two cases $8 \pi$ is included as
an amplification factor for the image). The distribution is
truncated when the orbits become unstable at $\mathcal{W}=3$, in
this case for $r \approx 0.633$. The curves labelled $N$ and $B$
are discussed below.}
\end{figure}

With the understanding that the procedure used here can be applied
to any rotation curve, we consider, as previously \cite{lake}, the
specific example of a universal curve given by
\cite{persicsalucci}
\begin{equation}
v(b)^2=\frac{\alpha b^2}{b^2+\beta},\label{ps}
\end{equation}
where $\alpha $ and $\beta$ are positive constants characteristic
of a particular galaxy. With (\ref{ps}) it follows form
(\ref{equivalence}) that
\begin{equation}
e^{2\Phi}=\frac{r^2}{\alpha
r^2\mathcal{W}(y)-\beta}\label{pspotentialsolution}
\end{equation}
where $\mathcal{W}$ is the Lambert W function \cite{lambert},
\begin{equation}
y = \frac{c e^{\frac{\beta}{\alpha r^2}}}{\alpha
^2r^2},\label{argument}
\end{equation}
and $c$ is a constant $>0$. Observe that with
(\ref{pspotentialsolution}) and (\ref{argument})
\begin{equation}
r \Phi^{'}=\frac{1}{1+\mathcal{W}},\label{w1}
\end{equation}
and so from (\ref{Phiprime}) the required angular momentum
distribution is given simply by
\begin{equation}
\mathcal{L}^2=\frac{1}{\mathcal{W}}.\label{w2}
\end{equation}
The gravitational mass follows from (\ref{M(r)}) and is given by
\begin{equation}
M=\frac{r}{3+\mathcal{W}}.\label{massdistribution}
\end{equation}
Finally,  from (\ref{mn}) or (\ref{mnl}) the number distribution
follows as
\begin{equation}
4\pi nm= \,{\frac {\mathcal{W }\left( \alpha\,{r}^{2} \left(
3+6\,\mathcal{W }+{\mathcal{W }}^{2} \right) +2\,
\beta\,\mathcal{W }\right) }{ \left( 3+\mathcal{W }\right) ^{2}
\left( 1+\mathcal{W }\right) ^{2}{r} ^{4}\alpha}}.\label{nw}
\end{equation}
In this model, mapping onto the observer's plane requires
$c>\alpha \beta$. With (\ref{w2}), limits on $\mathcal{W}$ follow
from the stability condition (\ref{stability}) and the existence
condition for the orbits so that
\begin{equation}
0<\mathcal{W}<3.\label{wlimits}
\end{equation}
Since $\mathcal{W}(y)$ is a monotone decreasing function of $r$,
decreasing from $\infty$ at $r=0$ to $0$ as $r\rightarrow\infty$,
condition (\ref{wlimits}) tells us that for stability the
distribution is necessarily a thick ``shell" with the central
divergence in  $\mathcal{W}(y)$  removed as mentioned above. The
junction conditions associated with this inner truncation are
discussed in detail below. Some representative properties of this
particular model are shown in Figure \ref{wmodelprop}

Because of the stability condition (\ref{stability}), a discussion
of junction conditions is appropriate. Geometrically, the junction
is examined by way of the Darmois-Israel conditions - the
continuity of the first and second fundamental forms at a boundary
surface $\Sigma$ \cite{musgrave}. For example, the smooth junction
of metrics of the form (\ref{standardform}), at any boundary
surface defined by constant $r$, only requires the continuity of
$M$ and $\Phi^{'}$, assuming the continuity of $r$, $\theta$ and
$\phi$. As a result, $G_r^r$, but not $G_{\theta}^{\theta}$ nor
$G_{t}^{t}$, is necessarily continuous at $r_{\Sigma}$. As a
consequence, since any Einstein cluster has $G_r^r=0$, the
matching conditions become trivial. One could, for example, insert
an interior or exterior Schwarzschild field at any $r_{\Sigma}$
simply by choosing the Schwarzschild mass $M$ to be
\begin{equation}
M=\frac{r^2
\Phi^{'}}{1+2r\Phi^{'}}|_{_{\Sigma}}.\label{Schwarzschild}
\end{equation}
This ``Swiss-cheese" type of matching is, however, clearly
inappropriate for the objects considered here \cite{schw}. Here
the distribution of ``dust" particles is supported by the hoop
stresses given by (\ref{thetatheta}) and junctions onto other
forms of ``dust" without these stresses are more appropriate in
the present context. Consider, as a simple example, junction onto
a uniform dust section (in particular, a piece of a flat
Robertson-Walker spacetime with density $\rho(t)$ \cite{rw} but
not necessarily the background cosmology). Let the junction
surface $\Sigma$ be defined by $M/r=\epsilon$ where $\epsilon$ is
a (dimensionless) constant. The continuity of the areal radii and
effective gravitational masses then gives
\begin{equation}
r_{\Sigma}=a(t)\textrm{r}_{\Sigma}=c\sqrt{\frac{\epsilon}{C}}a(t)^{3/2}
\label{sigmaevol}
\end{equation}
where $c$ is the velocity of light and $C$ is the constant defined
by $4 \pi \rho(t)  a(t)^3/3$. Unlike the ``Swiss-cheese" type of
model, here both $r_{\Sigma}$ and $\textrm{r}_{\Sigma}$ evolve
with $t$.

 It is interesting to compare the
model presented here with what would result from pure Newtonian
gravity. To do this consider the potential $\widetilde{\Phi}$ and
density $\tilde{\rho}$ defined by \cite{landl}
\begin{equation}
\nabla^2 \widetilde{\Phi}=4\pi \tilde{\rho} \equiv -R_{t}^{t}
\label{newtonpotential}
\end{equation}
where $R_{t}^{t}$ is the time component of the Ricci tensor of
(\ref{standardform}). A direct computation shows that
\begin{equation}
\tilde{\rho}=mn\left(\frac{1+r\Phi^{'}}{1-r\Phi^{'}}\right)
\label{newtonrho}
\end{equation}
where $mn$ is given by (\ref{mn}), or equivalently from
(\ref{Phiprime}),
\begin{equation}
\tilde{\rho}=mn(1+2\mathcal{L}^2). \label{newtonrhol}
\end{equation}
In pure Newtonian gravity we have
\begin{equation}
\rho=mn \label{newtonpure}
\end{equation}
and so pure Newtonian gravity fails to account for the
contribution made by the angular momentum distribution of the
orbiting particles.  This contribution is in fact the sole support
for the Einstein cluster.

We end with a few words about gravitational lensing. Of course any
model of a galactic halo should not only reproduce the observed
rotation curves, but also agree with any observed lensing. At
present the possibility of such simultaneous measurements seems
remote \cite{lensing}. Here we only demonstrate that an Einstein
cluster can be expected to be a significant source of
gravitational lensing. The curves $B$ and $N$ in Figure
\ref{wmodelprop} show the potential impact parameters
$\mathcal{B}$ \cite{lensingb}  for non-radial null geodesics in
the Newtonian case ($N$, no bending) and for the Einstein cluster
described above ($B$, for the values of $\alpha, \beta$ and $c$
stated). The flattening of $B$ is clear from
(\ref{massdistribution}).

\bigskip

In summary, we have presented a model, based on an argument
originally due to Einstein \cite{einstein}, that exactly
reproduces the rotation curve of any galaxy. The model requires no
new physics but does require the existence of WIMPs on stable
circular geodesic orbits. It is these orbiting particles that
create the gravitational field of the dark matter halo. The halo
is held together by the hoop stress generated by the angular
momentum of the particles. The actual value of the rest mass for
these particles plays no role in the argument which, given that
this is not presently know, can be considered an advantage. There
is no exact Newtonian counterpart to the model since pure
Newtonian gravity fails to account for the contribution made by
the angular momentum distribution of the particles in creating the
gravitational field. This contribution is in effect the entire
content of the model. The model predicts, on the basis of the
stability of orbits, that the halo is shell-like and so the WIMPs
do not occupy the inner regions, at least on stable circular
orbits. The gravitational lensing properties of the model need to
be compared with observations when these become available.  These
properties involve an analysis of the junction of the halos onto
both interior and exterior dust fields  which are not supported by
hoop stresses.

\begin{acknowledgments}
This work was supported by a grant from the Natural Sciences and
Engineering Research Council of Canada. Portions of this work were
made possible by use of \textit{GRTensorII} \cite{grt}.
\end{acknowledgments}


\begin{thebibliography}{}\label{sec:TeXbooks}
\bibitem[*]{email}{Electronic Address: lake@astro.queensu.ca}
\bibitem{einstein}A. Einstein, Ann. Math. \textbf{40}, 922 (1939).
Whereas Einstein was trying to show that the Schwarzschild horizon
can not exist since it does not exist in this very reasonable
model, this should not detract from the ingenious argument he
presented.
\bibitem{dark} The existence of dark matter is an integral part of
the model discussed here. In this regard see D. Garfinkle, Class.
Quant. Grav. \textbf{23}, 1391 (2006) {\tt arXiv:(gr-qc/0511082)}
and references therein.
\bibitem{heuristic} See G. L. Gomer, Gen. Rel. Grav. \textbf{28}, 601 (1996),
G. L. Comer, D. Langlois and P. Peter, Class. Quantum Grav.
\textbf{10}, L127 (1993), G. L. Comer and J. Katz, Class. Quantum
Grav. \textbf{10}, 1751 (1993) and A. B. Evans, Gen. Rel. Grav.
\textbf{8}, 155 (1977). This latter reference includes extensions
beyond the static case that we do not use here. See also, G.
Magli, Class. Quantum Grav. \textbf{14}, 1937 (1997),  A. F.
Teixeira and M. M. Som, J. Phys. A \textbf{7}, 838 (1974), H.
Bondi, Gen. Rel. Grav. \textbf{2}, 321 (1971), B. K. Datta, Gen.
Rel. Grav. \textbf{1}, 19 (1970), C. Gilbert, Mon. Not. R. Astr.
Soc. \textbf{114}, 628 (1954).
\bibitem{metric}We use geometrical units throughout, a signature of +2 and the  summation
convention. Functional dependence is shown usually only on the
first appearance of a function.
\bibitem{lake} K. Lake, Phys. Rev. Lett. \textbf{92}, 051101 (2004) {\tt
arXiv:(gr-qc/0302067)}. Refer to the references given there and
especially  U. Nucamendi, M. Salgado and D. Sudarsky, Phys. Rev. D
\textbf{63}, 125016 (2001) {\tt arXiv:(gr-qc/0011049)} on
extracting information from rotation curves. For a more recent
discussion, including the possible use of gravitational lensing,
see T. Faber and  M. Visser {\tt arXiv:(astro-ph/0512213)}.
\bibitem{persicsalucci} See M. Persic, P. Salucci and F. Stel, Mon. Not. R. Astr.
Soc. \textbf{281}, 27 (1996) {\tt arXiv:(astro-ph/9506004)}. On
the basis of a large number rotation curves, they have suggested a
universal rotation curve intrinsic to galactic halos as given in
(\ref{ps}). We continue to use it here as a demonstration to show
how observations determine $\mathcal{L}$ and therefore $mn$.
\bibitem{lambert}$\mathcal{W}$ is defined by the condition
$\mathcal{W}(x)e^{\mathcal{W}(x)}=x$ and the function has wide
application in the physical sciences. Despite the somewhat ugly
nature of $\mathcal{W}(y)$, the view taken here is that it is
simply an ``elementary function". See R. M. Corless, G. H. Gonnet,
D. E. G. Hare, D. J. Jeffrey, and D. E. Knuth, Advances in
Computational Mathematics \textbf{5}, 329 (1996).
\bibitem{musgrave}See, for example
P. Musgrave and K. Lake, Class. Quantum Grav. \textbf{13}, 1885
(1996). {\tt arXiv:(gr-qc/9510052)}
\bibitem{schw} It is interesting to note, however, that on the
small scale such a cluster around an object would make it appear
more relativistic on the outer boundary.
\bibitem{rw}We use $\textrm{r}$ as the comoving coordinate in the
uniform dust and $a(t)$ as the scale factor where $t$ is the
proper time for comoving observers.
\bibitem{landl} See, for example, L. D. Landau and E. M. Lifchitz, \textit{The Classical Theory of
Fields}, (Pergamon Press 1975) or S. Weinberg, \textit{Gravitation
and Cosmology: Principles and Applications of the General Theory
of Relativity}, (John Wiley and Sons, New York, 1972) or C. W.
Misner, K. S. Thorne and J. A. Wheeler, \textit{Gravitation}, (W.
H.  Freeman, San Francisco, 1973).
\bibitem{lensing} See, for example Faber and Visser \cite{lake}.
\bibitem{lensingb}From (\ref{standardform}) it follows that the evolution
of non-radial null geodesics is governed by
\begin{equation}
\left(\frac{dr}{d
\lambda}\right)^2=\frac{1}{r^2}\left(1-\frac{2M(r)}{r}\right)\left(\frac{\mathcal{B}(r)^2}{b^2}-1\right)\label{null}
\end{equation}
where $\lambda$ is an affine parameter, $b$ is a constant of
motion (the classical impact parameter at infinity in the
asymptotically flat case) and the potential impact parameter
$\mathcal{B}$ is given by
\begin{equation}
\mathcal{B}(r)^2=\frac{r^2}{e^{2\Phi(r)}}. \label{nullb}
\end{equation}
The straight line motion of pure Newtonian gravity is given by
$\mathcal{B}(r)=r$.
\bibitem{grt}This is a package which runs within Maple. It is entirely
distinct from packages distributed with Maple and must be obtained
independently. The GRTensorII software and documentation is
distributed freely on the World-Wide-Web from the address \textit{
http://grtensor.org}
\end{thebibliography}
\end{document}